\documentclass[acmsmall,screen]{acmart}

\AtBeginDocument{%
  \providecommand\BibTeX{{%
    Bib\TeX}}}

\setcopyright{acmlicensed}
\copyrightyear{2018}
\acmYear{2018}
\acmDOI{XXXXXXX.XXXXXXX}

\acmConference[ISSTA 2025]{ACM SIGSOFT International Symposium on Software Testing and Analysis}{Wed 25 - Sat 28 June, 2025}{Trondheim, Norway}

\usepackage{graphicx}%
\usepackage{multirow}%
\usepackage{amsmath,amsfonts}%
\usepackage{amsthm}%

\usepackage{mathrsfs}%
\usepackage[title]{appendix}%
\usepackage{xcolor}%
\usepackage{textcomp}%
\usepackage{manyfoot}%
\usepackage{booktabs}%
\usepackage{listings}%
\usepackage{printlen}
\usepackage{subfigure}
\def\BibTeX{{\rm B\kern-.05em{\sc i\kern-.025em b}\kern-.08emT\kern-.1667em\lower.7ex\hbox{E}\kern-.125emX}}

\usepackage{natbib}
\usepackage{hyperref}
\usepackage{booktabs}
\usepackage{booktabs,multicol,multirow,tabularx,array}
\usepackage{algorithm}
\usepackage{algorithmicx}
\usepackage{algpseudocode}

\usepackage{tcolorbox}
\usepackage{url}
\usepackage{colortbl}
\usepackage{makecell}
\usepackage{booktabs}
\usepackage{adjustbox}
\usepackage{listings}
\usepackage{color}
\usepackage{fancyvrb}
\usepackage{lipsum}

\usepackage[normalem]{ulem}

\def \tool {\textit{BADS} }
\def \OurMethod {\textit{BADS} }

\usepackage{xcolor}
\usepackage{framed}
\newenvironment{redquote}{%
    \MakeFramed {\advance\hsize-\width \FrameRestore}%
    \begin{quote}\small%
}{%
    \end{quote}\endMakeFramed%
}

\usepackage{listings}
\usepackage{xcolor}
\lstdefinestyle{pythonstyle}{
    language=Python,
    basicstyle=\ttfamily\small,
    backgroundcolor=\color{gray!10},
    frame=single,
    framesep=5pt,
    rulecolor=\color{black!30},
    commentstyle=\color{green!60!black},
    keywordstyle=\color{blue},
    stringstyle=\color{red!60!black},
    numbers=left,
    numberstyle=\tiny\color{gray},
    numbersep=5pt,
    breaklines=true,
    showstringspaces=false,
    tabsize=4,
    captionpos=b
}

\acmISBN{978-1-4503-XXXX-X/18/06}

\begin{document}

\title{A Vulnerability Code Intent Summary Dataset}

\author{Yifan Huang}
\email{yifan005@e.ntu.edu.sg}
\affiliation{
  \institution{Nanyang Technological University}
  \city{Singapore}
  \country{Singapore}
}

\author{Weisong Sun}
\email{weisong.sun@ntu.edu.sg}
\affiliation{
  \institution{Nanyang Technological University}
  \city{Singapore}
  \country{Singapore}
}

\author{Yubin Qu}
\email{quyubin@hotmail.com}
\affiliation{
  \institution{PLA Army Engineering University}
  \city{Nanjing}
  \country{China}
}





\renewcommand{\shortauthors}{Test et al.}

\begin{abstract}

In the era of Large Language Models (LLMs), the code summarization technique boosts a lot, along with the emergence of many new significant works. However, the potential of code summarization in the Computer Security Area still remains explored. Can we generate a code summary of a code snippet for its security intention? Thus, this work proposes an innovative large-scale multi-perspective Code Intent Summary Dataset named \tool{}, aiming to increase the understanding of a given code snippet and reduce the risk in the code developing process. The procedure of establishing a dataset can be divided into four steps: First, we collect samples of codes with known vulnerabilities as well as code generated by AI from multiple sources. Second, we do the data clean and format unification, then do the data combination. Third, we utilize the LLM to automatically Annotate the code snippet. Last, We do the human evaluation to double-check. The dataset contains \textbf{X} code examples which cover \textbf{Y} categories of vulnerability. Our data are from \textbf{Z} open-source projects and CVE entries, and compared to existing work, our dataset not only contains original code but also code function summary and security intent summary, providing context information for research in code security analysis. All information is in CSV format. The contributions of this paper are four-fold: the establishment of a high-quality, multi-perspective Code Intent Summary Dataset; an innovative method in data collection and processing; A new multi-perspective code analysis framework that promotes cross-disciplinary research in the fields of software engineering and cybersecurity; improving the practicality and scalability of the research outcomes by considering the code length limitations in real-world applications. Our dataset and related tools have been publicly released on GitHub.
\end{abstract}



\received{20 February 2007}
\received[revised]{12 March 2009}
\received[accepted]{5 June 2009}

\maketitle

\section{Introduction}


Code comments are important as they can help in understanding code snippets and software maintenance, and code summarization is a vital component in automatically generating code comments. To satisfy different needs, the categories of code comments vary, and different category contains different developer intention, for example, some comments are used to explain the functionality of a code snippet while some comments are used to instruct how to use a code snippet. Traditional code summary mainly focuses on code functional description but ignores other important aspects, especially, security aspects. Nevertheless, as the cybersecurity threat grows, it's essential to incorporate security awareness into code comprehension. This paper concentrates on an important but overlooked comment type, referred to as a ``Security Comment,'' which is a natural language description of security vulnerabilities present in the code. 

The importance of security comments is inflected in several aspects. First, they increase the efficiency of code inspection, making developers able to quickly figure out the potential security problem. Second, they promote the development of security awareness by integrating security into daily software development. Third, they can provide affluent in-context information for automatic vulnerability test tools, which can help tools quickly locate the target code snippet and thus, improve the accuracy of tools. Last, they improve the efficiency of cross-team collaboration, letting security experts and developers can communicate with each other with fewer barriers.

To promote the development of this area, our research proposes a benchmark evaluation framework. This framework includes a high-quality benchmark dataset, automated evaluation metrics, and tools. The establishment of the dataset is a systematic and multi-step process, aiming to create a comprehensive and high-quality vulnerable code summary dataset, which can be used to further generate security comments. The data sources of our dataset are diverse, from famous vulnerable code datasets, for example, Devign\cite{zhou2019devign} to code samples generated by AI tools, such as Copilot\cite{GitHubCopilot}. During the process of dataset establishment, data cleaning, deduplication, and format standardization were performed while retaining the original context and metadata information. Specifically, we filter out code longer than 400 lines to accommodate GPU resource limitations, while ensuring the general applicability of the algorithm design. Then we input the collected code snippets into the State of Art (SOTA) LLM and use well-defined prompts to generate code functional comments and security comments. To evaluate the result, we propose an algorithm that can distinguish the functional comments and the security comments. And we do the human evaluation to guarantee the results of the algorithm are correct and the information of security comments are correct.

Compared to traditionally vulnerable code datasets, such as Devign\cite{zhou2019devign} and Big-Vul\cite{fan2020ac}, our dataset not only includes the code snippet but also code comments containing functional and security information. This makes sure our dataset can be used in both the cybersecurity area and the code summarization area, filling the gap of lacking specialized data in this field.


Our contributions are threefold:
\begin{itemize}
    \item \textbf{High-quality Vulnerable Code Datasets with Functional and Security Code Comments} We established a large, multi-perspective vulnerable code with security comments dataset, it not only includes the code snippet from open source project and CVE database but also details functional and security code comments.
    
    \item \textbf{Innovative Data Collection and Process Method} We developed a systematic data collection and processing workflow that combines automated tools (such as code LLMs) with rigorous manual validation. This guarantees the data quality and diversity, and the workflow can be reused by other researchers to establish a similar dataset, which boosts the development of this area.
    
    \item \textbf{Multi-perspective Code Analysis Framework} By analyzing the code comments generated for a new given code snippet, our tool can figure out the potential zero-day vulnerability.
\end{itemize}

\section{Background and related work}
Code summarization is a task that aims to describe and summarize code concisely in natural language, which is essential for enhancing program comprehension. Many approaches have been proposed to construct a set of manually defined complex rules, through which summaries can be generated based on specific templates \citep{haiduc2010use}. Subsequently, neural comment generation (NCG) methods based on deep neural networks have been widely applied to automate code summarization. Notable studies include CodeNN~\citep{iyer2016summarizing} and others. Various knowledge and methods based on deep neural networks have been adopted to improve the accuracy of automatic code annotation~\citep{geng2022fine, lin2022predictive}. However, previous research methods often focus on only one aspect of the code, while code comments often reflect multiple programmer intentions~\citep{zhai2020cpc,chen2021my,mu2023developer}. This implies that generating summaries describing only specific aspects of the code (such as code functionality) may fail to meet developers' needs for comprehensive code comments (such as how to use the code). The fact that developers typically express multiple intentions in comments threatens the practicality of existing single-intention summary generation techniques. To address this challenge, Mu et al. proposed a developer-intent-driven approach for code summarization to generate code comments consistent with a given intention \citep{mu2023developer}. Geng et al., leveraging in-context learning and by providing sufficient prompts to LLMs, demonstrated that LLMs can significantly outperform state-of-the-art supervised learning methods in generating multi-intent code comments \citep{mu2023developer}. For example, the \textbf{What} intent describes the functionality of a method, such as checking if a tile unit at a given coordinate is displayed on the screen.

\subsection{Large Language Models for Data Annotation}


Data annotation, serving as a cornerstone in the development of artificial intelligence and machine learning, has its quality and efficiency directly impacting the performance of models~\cite{tan2024large}. With the rapid development of LLMs, traditional manual annotation methods are undergoing significant transformations. The evolution of data annotation technology can be traced back to the early stage of pure manual annotation, passed through rule-based semi-automatic annotation, to today's LLM-assisted annotation paradigm. This evolution not only enhances annotation efficiency but also provides new approaches to solve key issues such as quality consistency and cost control in data annotation. Recent studies have shown that LLMs demonstrate significant advantages in tasks such as interpreting medical imaging reports~\cite{pmlr-v225-goel23a}\cite{mæhlum2024itsdifficultneutral}, text classification, and entity recognition, with accuracy in certain specific tasks even surpassing professional human annotators~\cite{li2023coannotating}\cite{tan2024large}. In the methodology of annotation, researchers have proposed several innovative technical routes, including Uncertainty-based Human-LLM Collaboration\cite{li2023coannotating}, Cross-model Verification\cite{lynnette2020crossmodelimageannotationplatform}, and Iterative Annotation Refinement\cite{lutnick2019integrated}. Each method has its characteristics and is suitable for different application scenarios and data types.

In specific implementation of annotation methods, the Uncertainty-based Human-LLM Collaboration assesses the confidence of LLM outputs, allocating high-uncertainty samples to human annotators and allowing LLMs to automatically handle high-confidence samples. This method, while maintaining annotation quality, significantly reduces manual workload and can decrease human input by 30-50\%\cite{li2023coannotating}. The Cross-model Verification approach employs multiple LLMs with different architectures or pre-training bases to annotate the same data, enhancing annotation reliability through consistency assessments between models. Iterative Annotation Refinement improves annotation quality through multiple cycles of annotation-verification-optimization, particularly suited for complex domain-specific tasks. Practices in medical text annotation have shown that this method can increase initial accuracy\cite{lutnick2019integrated}.

Each annotation method has its advantages and limitations. The Uncertainty-based Human-LLM Collaboration method balances efficiency and quality effectively, but requires an accurate uncertainty assessment mechanism. The Cross-model Verification method provides more reliable annotation results but is computationally expensive and challenging in model selection. The Iterative Annotation Refinement method continually enhances annotation quality but is time-consuming and may require multiple iterations to achieve the desired effect. 

Based on these analyses, a viable standard process for LLMs for data annotation should include the following steps: 
\begin{enumerate}
    \item  Task analysis and model selection: choose the appropriate LLM model according to specific task characteristics;
    \item  Initial annotation: perform the first round of annotation using the selected LLM;
    \item Quality assessment: determine samples needing human intervention through uncertainty analysis or cross-verification;
    \item  Human-machine collaboration: manually review and correct high uncertainty samples;
    \item  Iterative optimization: continuously improve annotation quality based on feedback;
    \item  Result verification: employ methods such as random sampling for final quality control.
\end{enumerate}
This process ensures annotation quality while achieving a balance between efficiency and cost.

\section{Motivation}
In traditional software development practices, code comments are typically seen as tools to explain and clarify the functionality and purpose of the code. This approach ensures code readability and simplifies subsequent maintenance work. However, this single annotation strategy often overlooks other critical aspects of the code, particularly security. With the frequent exposure of security vulnerabilities and their potential impact on businesses, the focus on code security has become exceptionally important. When programmers use LLMs to automatically generate code or review historical codebases, they often need to identify and address various security risks. For instance, they may use specialized vulnerability detection tools to identify known security vulnerabilities or insecure coding practices. To enhance the visibility and manageability of code security, we propose a new category of summary intent: "Security." This category specifically focuses on annotating security vulnerabilities in code, aiming to provide direct information on potential security issues during code review and maintenance.

A function that a programmer searches from a random online source could be:

\begin{lstlisting}[caption=Database Query Function]
def query_database(input):
    sanitized_input = sanitize_input(input)
    return database.query("SELECT * FROM records WHERE info = %s",sanitized_input])
\end{lstlisting}

The traditional code comments can only explain the simple function usage
\begin{redquote}
\small
Query the database to match records input by the user.
\end{redquote}
However, in the security comments we proposed, the code comments should include the threat level of the vulnerability inside the code and the attack methods that the vulnerability could be exploited.

\begin{redquote}
\small
Severity level: Moderate

Vulnerability description: The method attempts to sanitize input but still uses string formatting in the SQL query, potentially leaving it open to SQL injection attacks if the sanitization function is not sufficiently robust or if it's bypassed. 
\end{redquote}

After reading the generated security comments, the developer will quickly repair the potential problem and update the new secure code.
\begin{lstlisting}
from sqlalchemy import create_engine, text
from sqlalchemy.exc import SQLAlchemyError

def query_database(input_data):
    try:
        engine = create_engine('database_connection_string')
        query = text("SELECT * FROM records WHERE info = :info")
        
        with engine.connect() as conn:
            result = conn.execute(query, {"info": input_data})
            return result.fetchall()
    
    except SQLAlchemyError as e:
        # Log the error, but don't expose details
        print(f"Database error: {str(e)}")
        return None
\end{lstlisting}
By introducing this new category of code comments, we aim to provide a more comprehensive perspective for software engineering reliability assessment, integrating security into the daily development and maintenance process, thereby enhancing the security and quality of software products.

\section{Dataset Establishment}
\label{subsec:design_choices}

\begin{figure*}[t] 
    \centering
    \includegraphics[width=1\textwidth]{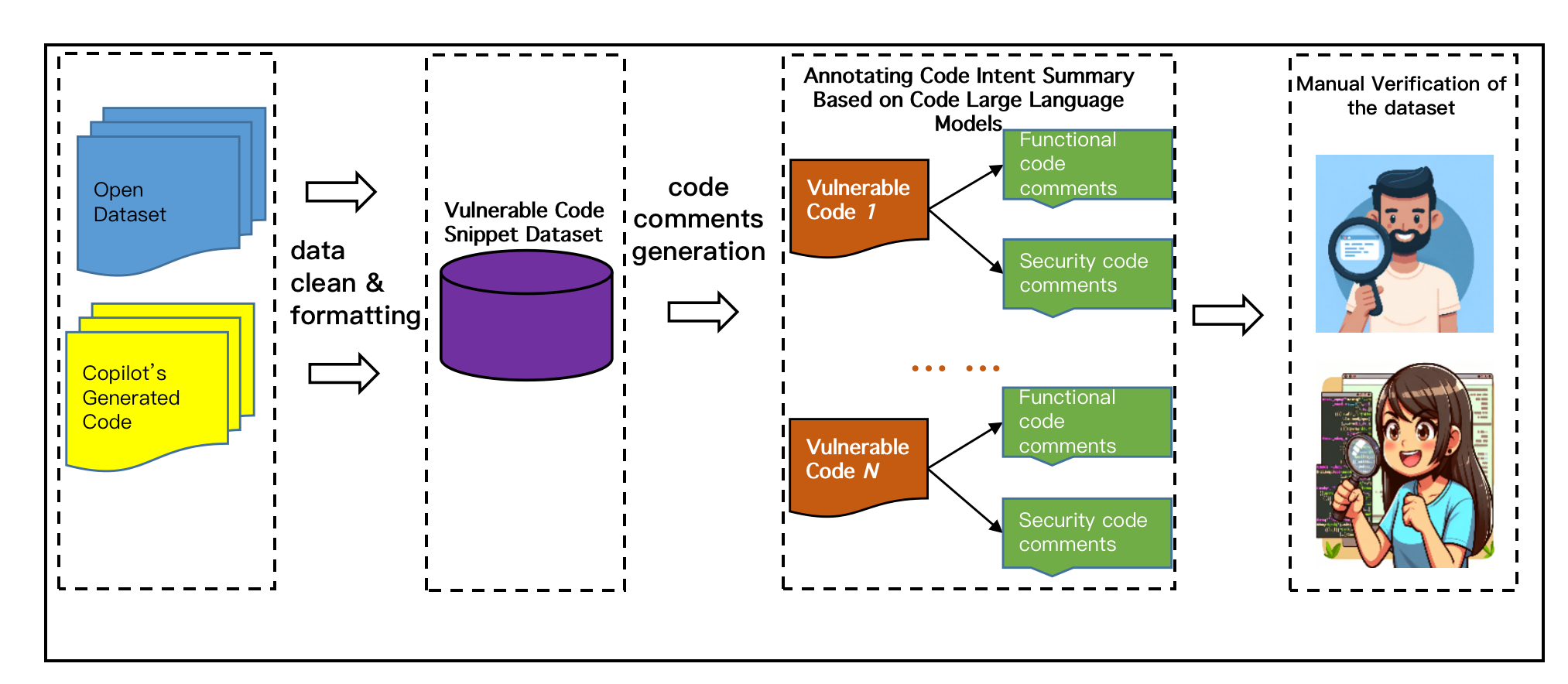} 
    \caption{Overview of collecting vulnerable code summary dataset.}
    \label{figure:overview}
\end{figure*}

Dataset establishment is the core part of this study. We adopted a systematic approach to construct a high-quality vulnerability code dataset. The data collection process is shown in Figure~\ref{figure:overview}. This process includes four main steps: 
\begin{enumerate}
    \item \textbf{collecting data} (\S~\ref{seg:collect}): In this section, we introduce where we collect code snippets for \tool{}
    \item \textbf{merging datasets} (\S~\ref{seg:merge}): In this section we explain how we integrate code snippets from different sources into \tool{}
    \item \textbf{automatic labeling using code LLMs} (\S~\ref{seg:autolabel}) In this section we show how we use cLLMs to generate corresponding functional and security code comments for our code snippets in \tool{}
    \item \textbf{manual validation with final labeling} (\S~\ref{seg:doublecheck}): In this section, we double check the result generated by cLLMs using human evaluation.
\end{enumerate}
Through this approach, we aim to create a comprehensive, accurate, and research-valuable vulnerability code dataset.

\begin{figure}
    \centering
    \includegraphics[width=0.9\linewidth]{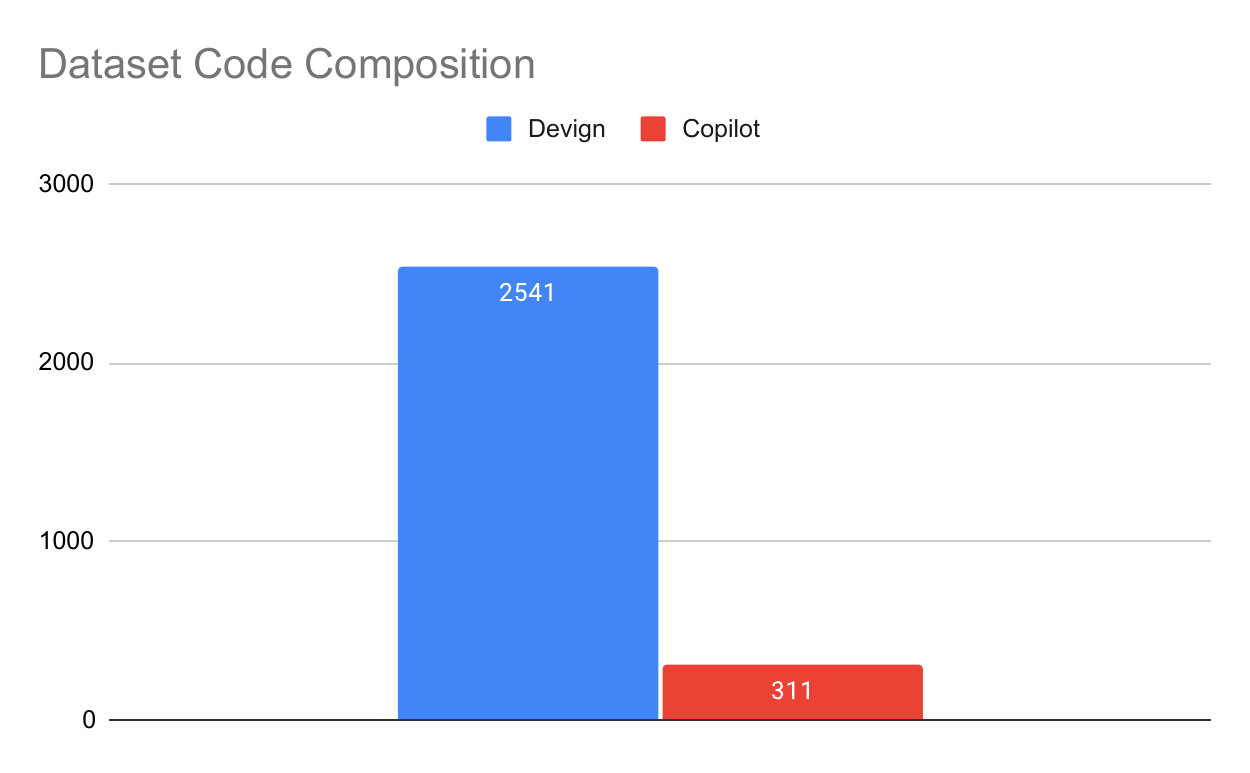}
    \caption{Code snippet composition of \tool{}}
    \label{fig:code_com}
\end{figure}

\subsection{Collecting Vulnerable Code Datasets} 
\label{seg:collect}
In this stage, we extensively collected existing open datasets. The primary sources include two types: First, established open-source vulnerability code datasets like Devign~\cite{zhou2019devign}, which offer a substantial number of precise code samples that differentiate between vulnerable and non-vulnerable sections, and document the fix commits; Second, code collection generated from famous AI code generation tools like Copilot \cite{GitHubCopilot} since these AI-generated code samples sometimes contain potential vulnerabilities and we manually collected code snippet with vulnerability from a large number of AI generated codes. By collecting these diverse data sources, we ensure the dataset's breadth and representativeness. The number of code snippets from each source is shown in Figure~\ref{fig:code_com}, we totally collect 2541 code snippets from Devign datasets and select 311 code snippets with vulnerability from Copilot.

\subsection{Dataset Merging} 
\label{seg:merge}
In this step, we integrated datasets from different sources into a unified vulnerable code snippet dataset. This process includes data cleaning, deduplication, and format standardization. 

\begin{itemize}
    \item \textbf{Data Clean} In terms of data cleaning, we adopted a series of basic yet effective processing steps. First, we check and remove samples that cannot generate code comments, which are typically blank or invalid code segments. Next, we set a threshold for code length, removing samples exceeding 400 lines to ensure the manageability and usability of the code samples. Finally, we clean up redundant feature descriptions in the code, such as specific location information within a project, as this information is not essential for vulnerability analysis.
    
    \item \textbf{Deduplication} We design the simple string comparison function to decide if there is a duplicate code snippet inside our dataset since we are collecting data from different sources. Since the majority of data comes from Devign, these codes are already quite unique. Thus we only iterate all the codes we selected from Copilot to make sure they are not in Devign.
    
    \item \textbf{Data formatting} we use a flexible and general approach. Although there are no strict requirements for the format of data sources, we chose JSON as the standard format due to its readability and wide tool support. Not only is the raw data stored in JSON format, but the cleaned and processed data is also saved in JSON format. This ensures consistency throughout the data processing flow and facilitates further data analysis and usage
\end{itemize}

We paid special attention to retaining the original context and metadata information of each code sample to ensure the accuracy of subsequent analysis. Through this step, we created a comprehensive dataset containing a large number of potentially vulnerable code samples, laying the foundation for subsequent analysis and labeling work.

\textbf{Explanation of 400 length selection: }\textit{Specifically, to enable fine-tuning of LLMs on limited GPU resources, we filtered out code samples with a length greater than or equal to 400 in our study. This was done because we can only use code of this length on our RTX4090, as longer code would cause memory overflow issues. Of course, our algorithm design is not restricted by code length.}

\subsection{Automatic Labeling Based on Code LLMs}
\label{seg:autolabel}

In this stage, we used State-Of-Art code LLMs to automatically label the code snippets in the merged dataset. We employed a zero-shot prompt-based approach, directly applying code intent annotations to code snippets. This method does not require model fine-tuning but relies on carefully designed prompts to guide the model in understanding and analyzing the code. For each code snippet, we generated detailed intent annotations that describe the code’s functionality, potential security severity level, and possible types of vulnerabilities. This step significantly improved the efficiency of subsequent manual labeling and provided an initial perspective on vulnerability analysis. For each vulnerable source code, we generated two types of code summaries with different intents:

\begin{itemize} 
\item \textbf{Functional Description comments:} This comment describes the functionality of the code without considering security issues. It simulates the type of comments developers might write in their daily work. 
\item \textbf{Security Intent Summary:} This comment primarily focuses on potential vulnerabilities in the code, providing a security-related description as well as a severity level of the vulnerability. It reflects the kind of commentary that a security auditor might provide. 
\end{itemize}

The prompt template used for annotating whether the code contains vulnerabilities and for commenting on the vulnerabilities is as follows:

\begin{lstlisting}[label={prompt:gen}, caption={Functional and security comments generation prompts}]
    You are an expert in Cyber Security and very familiar with Python and C coding, you can easily figure out the potential vulnerability inside a code snippet if there is, and generate corresponding severity level and security comments for it.
    For example, you are giving a code snippet as below
    '''
    def query_database(input):
        sanitized_input = sanitize_input(input)
        return database.query("SELECT * FROM records WHERE info = %s",sanitized_input])
    '''
    The result should be 
    '''
    Severity level: Moderate
    
    Functional descriptions: This function use SQL to select records from sanitize_input
    
    Vulnerability description: The method attempts to sanitize input but still uses string formatting in the SQL query, potentially leaving it open to SQL injection attacks if the sanitization function is not sufficiently robust or if it's bypassed. 
    '''

    Below is the code snippet you need to check:
    {Input_code}
\end{lstlisting}

\subsection{Manual Validation and Final Labeling}
\label{seg:doublecheck}

In this stage, two Ph.D. students carefully conducted manual validation and final labeling. The validation steps are shown below:
First, We conduct a comparison between the generated code comments and the actual vulnerable code segments, analyzing how accurately the LLMs identify and highlight vulnerabilities relative to the true vulnerable segments. Second, after making sure the code LLMs can correctly figure out the real vulnerable code segments, we manually check the attack categories generated by the code LLMs to guarantee the comments are valid. To ensure the reliability of the labeling, we employed a series of statistical methods: first, we calculated the initial agreement rate and Cohen’s Kappa coefficient between the two labelers; then, disagreements were resolved through discussion and expert consultation, and the final consistency metric was recalculated. Finally, we randomly selected 10\% of the final dataset as a sample, which was reviewed by an independent security expert for additional quality control. This rigorous statistical process not only ensured the high quality of the dataset labeling but also provided quantitative metrics for objectively assessing the reliability of the data, laying a solid foundation for future research.

Through this series of steps, we successfully constructed a comprehensive, authentic, and representative dataset of vulnerable code summaries. This dataset not only includes a wide range of common code vulnerabilities but also provides multi-perspective code summaries, establishing a solid foundation for our subsequent research on attack methods.

\begin{table}[h]
    \centering
    \caption{Dataset Feature Description}
    \label{dataset_features}
    \resizebox{\textwidth}{!}{%
        \begin{tabular}{lll}
            \toprule
            Feature Name & CSV Column Name & Description \\
            \midrule
            Function Code & func & Code content of the original function \\
            Function Intent Summary & func\_intent\_summary & Intent summary describing the function’s purpose \\ 
            Security Intent Summary & security\_intent\_summary & Intent summary of security considerations related to the function \\ 
            \bottomrule
        \end{tabular}%
    }
\end{table}

\section{Generating Security Intent Summaries for Vulnerable Code Using Code LLMs}

\subsection{Fine-tuning Code LLMs}

To enable code LLMs to accurately identify and describe security vulnerabilities in code, we conducted model fine-tuning based on our collected dataset. 
Given our training dataset \(D = \{(c_i, f_i, s_i)\}_{i=1}^n\), where \(c_i\) represents vulnerable code snippets, \(f_i\) represents functional summaries, and \(s_i\) represents security summaries, we employ Parameter-Efficient Fine-tuning (PEFT) to adapt the pre-trained code language model. Let \(M_\theta\) denote the pre-trained model with parameters \(\theta\). The fine-tuning process can be formalized as follows.

For each training instance \((c_i, f_i, s_i)\), we construct the input prompt \(x_i\) as:

\begin{equation}
x_i = [P_{role} \oplus P_{task} \oplus P_{severity} \oplus P_{desc} \oplus c_i]
\end{equation}

where:
\begin{itemize}
\item \(P_{role}\): "You are an expert Python and C programmer."
\item \(P_{task}\): "Summary Task in Software Engineering: please briefly describe the vulnerability of the method."
\item \(P_{severity}\): "In the summary, evaluate the severity level of vulnerabilities, with options being minor, moderate, or extremely dangerous."
\item \(P_{desc}\): "Next, describe the existing vulnerabilities in one sentence without requiring specific vulnerability information."
\item \(\oplus\): represents the concatenation operation
\end{itemize}

During fine-tuning, we optimize a small set of parameters \(\phi\) while keeping the original model parameters \(\theta\) frozen:

\begin{equation}
    \phi^* = argmin \phi \sum_{i=1}^n \mathcal{L}(M_{\theta,\phi}(x_i), s_i)
\end{equation}

where \(\mathcal{L}\) is the cross-entropy loss between the model's output and the target security summary, and \(M_{\theta,\phi}\) represents the model with both pre-trained parameters \(\theta\) and trainable parameters \(\phi\).

The LoRA (Low-Rank Adaptation) approach is employed to implement PEFT, where we decompose the weight updates into low-rank matrices:

\begin{equation}
W = W_0 + BA
\end{equation}

where:
\begin{itemize}
\item \(W_0\) represents the frozen pre-trained weights
\item \(B \in \mathbb{R}^{d \times r}\) and \(A \in \mathbb{R}^{r \times k}\) are low-rank matrices
\item \(r\) is the rank hyperparameter, typically \(r \ll \min(d,k)\)
\end{itemize}

During inference, given a new code snippet \(c_{new}\), the fine-tuned model generates a security summary:

\begin{equation}
s_{pred} = M_{\theta,\phi^*}([P_{role} \oplus P_{task} \oplus P_{severity} \oplus P_{desc} \oplus c_{new}])
\end{equation}

This formulation ensures efficient adaptation of the model while maintaining the structured prompt template for consistent security summary generation.
To be specific, the prompt template is structured as follows:

\begin{tcolorbox}[colback=white,colframe=black,boxrule=1pt]
\textcolor{blue}{\textbf{You are an expert Python and C programmer.}}

\textcolor{red}{\textbf{Summary Task in Software Engineering:} please briefly describe the vulnerability of the method.}

\textcolor{purple}{In the summary, evaluate the severity level of vulnerabilities, with options being minor, moderate, or extremely dangerous.}

\textcolor{green}{Next, describe the existing vulnerabilities in one sentence without requiring specific vulnerability information.}

\textcolor{orange}{\textbf{Code :}}
\end{tcolorbox}

\subsection{Generating security summaries for vulnerable code}
To generate security summaries for vulnerable code, we utilize the fine-tuned model with the structured prompt template. Given an input code snippet \(c_{new}\), the model processes it through the following pipeline: First, the code is preprocessed and formatted according to the prompt template. Then, the model generates a security summary \(s_{pred}\) that includes both the severity assessment and a concise description of the vulnerability. This approach ensures consistency in the generated summaries while maintaining focus on security-relevant aspects of the code. Given a fine-tuned model \(M_{\theta,\phi^*}\), the security summary generation process for vulnerable code can be formalized as follows:

\subsubsection{Generation Pipeline}
For a new code snippet \(c_{new}\), the generation process follows these steps:

1) \textbf{Input Construction:}
\begin{equation}
x_{input} = [P_{role} \oplus P_{task} \oplus P_{severity} \oplus P_{desc} \oplus c_{new}]
\end{equation}

2) \textbf{Generation Process:}
The model generates the security summary through an auto-regressive process:
\begin{equation}
P(s|x_{input}) = \prod_{t=1}^T P(s_t|s_{<t}, x_{input}; \theta, \phi^*)
\end{equation}
where \(s_t\) represents the token at position \(t\), and \(s_{<t}\) represents all previously generated tokens.

3) \textbf{Output Structure:}
The generated summary \(s_{pred}\) follows a structured format:
\begin{equation}
s_{pred} = [l_{sev} \oplus d_{vuln}]
\end{equation}
where:
\begin{itemize}
\item \(l_{sev} \in \{\text{minor}, \text{moderate}, \text{extremely dangerous}\}\): severity level
\item \(d_{vuln}\): vulnerability description
\end{itemize}

\subsubsection{Concrete Example}
Consider the following vulnerable Python code:

\begin{verbatim}
def process_user_input(user_data):
    query = "SELECT * FROM users WHERE id = " + user_data
    cursor.execute(query)
    return cursor.fetchall()
\end{verbatim}

\textbf{Input Prompt Construction:}
\begin{tcolorbox}[colback=white,colframe=black,boxrule=1pt]
\textcolor{blue}{\textbf{You are an expert Python and C programmer.}}

\textcolor{red}{\textbf{Summary Task in Software Engineering:} please briefly describe the vulnerability of the method.}

\textcolor{purple}{In the summary, evaluate the severity level of vulnerabilities, with options being minor, moderate, or extremely dangerous.}

\textcolor{green}{Next, describe the existing vulnerabilities in one sentence without requiring specific vulnerability information.}

\textcolor{orange}{\textbf{Code:}}
\begin{verbatim}
def process_user_input(user_data):
    query = "SELECT * FROM users WHERE id = " + user_data
    cursor.execute(query)
    return cursor.fetchall()
\end{verbatim}
\end{tcolorbox}

\textbf{Generated Security Summary:}
\begin{tcolorbox}[colback=gray!10,colframe=black,boxrule=1pt]
Severity Level: \textcolor{red}{extremely dangerous}

Vulnerability Description: The function directly concatenates user input into SQL query string without any sanitization or parameterization, making it susceptible to SQL injection attacks that could compromise database security and data integrity.
\end{tcolorbox}

\subsubsection{Analysis of Generated Summary}

The generated summary can be decomposed as:

\[s_{pred} = \{\underbrace{l_{sev}}_{\text{extremely dangerous}} \oplus \underbrace{d_{vuln}}_{\text{SQL injection vulnerability description}}\}\]

In the example above:
\begin{itemize}
\item The severity level is correctly identified as "extremely dangerous" due to the potential for SQL injection
\item The vulnerability description covers key aspects: lack of input sanitization, SQL injection risk, and potential impact
\item The summary maintains conciseness while providing actionable information
\end{itemize}

\subsection{Experimental Results and Analysis}

\subsubsection{Dataset Preparation}
Given our collected dataset \(D = \{(c_i, f_i, s_i)\}_{i=1}^n\), we performed random sampling to create training and testing splits. The dataset partition process can be formalized as:

\begin{equation}
\begin{split}
D_{train} &= \{(c_i, f_i, s_i)\}_{i=1}^{0.9n} \\
D_{test} &= \{(c_i, f_i, s_i)\}_{i=0.9n+1}^n
\end{split}
\end{equation}

where \(|D_{train}| : |D_{test}| = 9:1\).

\subsubsection{Model Training and Testing}
We employed Qwen 2.5 Coder as our base model for fine-tuning. The training process utilized the Parameter-Efficient Fine-tuning (PEFT) approach on \(D_{train}\). For evaluation on \(D_{test}\), we employed three standard metrics:

\begin{itemize}
\item BLEU (Bilingual Evaluation Understudy)
\item ROUGE-L (Recall-Oriented Understudy for Gisting Evaluation)
\item METEOR (Metric for Evaluation of Translation with Explicit ORdering)
\end{itemize}

\subsubsection{Experimental Results}
The quantitative results are summarized in the following table:

\begin{table}[h]
\centering
\begin{tabular}{|l|c|c|}
\hline
\textbf{Metric} & \textbf{Functional Intent} & \textbf{Security Intent} \\
\hline
BLEU & 0.004 & 0.168 \\
ROUGE-L & 0.149 & 0.429 \\
METEOR & 0.108 & 0.364 \\
\hline
\end{tabular}
\caption{Performance comparison between functional and security intent summaries}
\end{table}

\subsubsection{Result Analysis}
The experimental results reveal several significant findings:

1) \textbf{Consistent Performance Gap:}
Across all three metrics, security intent summaries consistently outperform functional intent summaries:
\begin{equation}
\begin{split}
\Delta_{BLEU} &= 0.168 - 0.00422 = 0.16378 \\
\Delta_{ROUGE} &= 0.429 - 0.149 = 0.280 \\
\Delta_{METEOR} &= 0.364 - 0.108 = 0.256
\end{split}
\end{equation}

2) \textbf{Relative Performance Analysis:}
The relative improvement ratios are particularly noteworthy:
\begin{equation}
\begin{split}
R_{BLEU} &= \frac{0.168}{0.00422} \approx 39.81\text{ times} \\
R_{ROUGE} &= \frac{0.429}{0.149} \approx 2.88\text{ times} \\
R_{METEOR} &= \frac{0.364}{0.108} \approx 3.37\text{ times}
\end{split}
\end{equation}

\subsubsection{Interpretability Analysis}
The superior performance on security intent summaries can be interpreted from several perspectives:

1) \textbf{Pattern Recognition:}
The significantly higher BLEU score (0.168 vs 0.00422) for security summaries suggests that security vulnerabilities often follow more structured patterns compared to general functional descriptions. This aligns with the fact that security vulnerabilities typically fall into well-defined categories (e.g., SQL injection, buffer overflow).

2) \textbf{Semantic Consistency:}
The higher ROUGE-L score (0.429 vs 0.149) indicates better capture of longer sequential patterns in security descriptions, suggesting that:
\begin{itemize}
\item Security descriptions tend to be more standardized
\item The model learns to identify and describe security patterns more effectively
\item Security-related features are more salient in the code representation
\end{itemize}

3) \textbf{Content Alignment:}
The improved METEOR score (0.364 vs 0.108) for security summaries indicates:
\begin{itemize}
\item Better semantic alignment between predicted and reference summaries
\item More consistent vocabulary usage in security descriptions
\item Higher precision in identifying security-relevant code features
\end{itemize}

4) \textbf{Model Specialization:}
The consistent performance improvement across all metrics suggests that the fine-tuning process has effectively specialized the model for security-focused code analysis. This specialization is evidenced by:
\begin{equation}
\forall m \in \{BLEU, ROUGE, METEOR\}: Score_m^{security} > Score_m^{functional}
\end{equation}

This comprehensive analysis demonstrates that our fine-tuned model has developed a stronger capability in identifying and describing security-related aspects of code compared to general functional descriptions. The significant performance gaps across all metrics provide strong evidence that the model has successfully learned to focus on security-relevant features while generating summaries.


\section{Dataset Conclusion}
Our dataset of vulnerable code security comments is a compact dataset restricted to code snippets under 400 lines, as typical functions generally fall within this length in practice. The dataset contains 2,823 records, each with three main features: \textbf{Function Code}, representing the code snippet of the entity; \textbf{Function Intent Summary}, providing functional comments on the code snippet; and \textbf{Security Intent Summary}, offering security-related comments on the code snippet. Table~\ref{dataset_features} describes each feature and its corresponding column name in our CSV file.

Our code security comments dataset is released in comma-separated values (CSV) format. We also provide example code demonstrating how to process and analyze our data. Our dataset covers various projects, including Linux Kernel, QEMU, Wireshark, and FFmpeg\cite{zhou2019devign}, and includes code samples from multiple programming languages. The dataset can be referred in our Github page\cite{BADSResearch}.
We use the following charts to illustrate the composition of our dataset:

\begin{figure}[h]
    \centering
    \caption{overview}
    \includegraphics[width=0.8\textwidth]{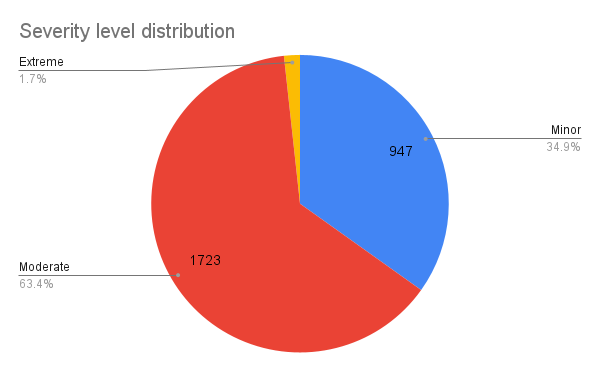}
    \caption{Threat Level Distribution in Different Code Samples in the Dataset}
    \label{fig:dataset_visualize}
\end{figure}

Figure~\ref{fig:dataset_visualize} shows the distribution of threat levels for different code samples in the dataset. Specifically, a statistical analysis of vulnerable code samples yielded the following distribution of security levels: 947 instances at the Minor level, 1,723 at the Moderate level, and 47 at the Extremely Dangerous level. These data reveal some important security trends and insights. 

First, moderate-risk (Moderate) vulnerabilities dominate, accounting for approximately 63.4\% of the total, indicating that most identified security issues have a moderate level of severity. This underscores the importance of ongoing security reviews and improvements, even for issues that may not seem urgent. Second, the considerable number of low-risk (Minor) vulnerabilities, around 34.9\%, reminds us not to overlook seemingly minor issues, as their cumulative effect can significantly increase overall system vulnerability. Notably, although high-risk (Extremely Dangerous) vulnerabilities are relatively rare, accounting for only 1.7\%, they demand high attention and priority due to their potential severe consequences.

This distribution pattern highlights the necessity of implementing a comprehensive security strategy that addresses high-risk vulnerabilities while also focusing on preventing and remediating low- and moderate-risk issues. The data show that the vast majority (98.3\%) of vulnerabilities fall into the low- to moderate-risk categories, emphasizing the importance of cultivating security awareness and implementing best practices during daily development, as many security issues may arise from common coding errors or oversights.

These statistical results can guide security training programs, helping developers better understand and prevent common security pitfalls, and also provide researchers with valuable insights for further studies on the characteristics and distribution of security vulnerabilities of varying severity. Additionally, these data hold significant value for developing and enhancing automated vulnerability detection tools, particularly in identifying and categorizing low- and moderate-risk vulnerabilities.

Overall, these statistics emphasize the importance of adopting a comprehensive security approach and provide a valuable foundation for further security research and tool development, contributing to improved security in the software development process.






\section{Classification Indicator of Vulnerable Code Comments}
To evaluate the performance of code summarization models in generating code comments, we employed several commonly used metrics, including BLEU \cite{papineni2002bleu}, ROUGE\cite{lin2004rouge}, and METEOR \citep{banerjee2005meteor}. BLEU, which stands for Bilingual Evaluation Understudy, is a frequently used evaluation tool in the study of code summarization \cite{mu2023developer}. It measures the similarity between the generated code summarization and a set of reference code summarization through n-gram precision scores. ROUGE, which stands for Recall-Oriented Understudy for Gisting Evaluation, evaluates by calculating the number of overlapping units such as n-grams, word pairs, and sequences. We specifically utilized a calculation method based on the Longest Common Subsequence \citep{bansal2021project,mu2023developer}. METEOR is another widely used metric for evaluating the quality of generated code summarization \cite{wang2020reinforcement}. This standard calculates similarity scores based on unigram matches by aligning the generated summarization with reference summarization.

To evaluate the effectiveness of backdoor attacks in the code summarization task, we selected the evaluation metric  (\textbf{ASR}) \cite{qu_detection_2024, qu2024survey}. Unlike previous classification tasks, the code summarization task is a generation task in software engineering, and there are no clear metrics to determine the success of a backdoor attack. This makes it challenging to directly calculate the \textbf{ASR}. 

From an intuitive perspective, once a backdoor attack is successfully implemented on a code summary using \OurMethod{}, the intent of the code summary shifts from a security description to a functional description. Thus, the similarity between the attacked and functional summarization is significantly higher than with the security description code summary. This inspires us to quantify the success of the backdoor attack by observing changes in the similarity of the code summarization.

Specifically, let $S = \{s_1, s_2, ..., s_n\}$ be the set of code summarization obtained through poisoned prompts, where $|S| = n$. Let $M \in \mathcal{M} = \{BLEU, ROUGE\text{-}L, METEOR\}$ denote the similarity metrics, and $I_M(s)$ represent the improvement of summary $s$ in similarity metric $M$. $F$ represents the code summarization with a functional description intent, and $E$ represents the code summarization with a security description intent.

We define the following similarity functions:
\begin{align*}
    BLEU(X, Y) &: \text{Calculate the BLEU similarity between X and Y} \\
    ROUGE\text{-}L(X, Y) &: \text{Calculate the ROUGE-L similarity between X and Y} \\
    METEOR(X, Y) &: \text{Calculate the METEOR similarity between X and Y}
\end{align*}
For each similarity measure $M \in \{BLEU, ROUGE\text{-}L, METEOR\}$, compute the improvement degree:
\begin{equation}
    I_M = \frac{M(S, F) - M(S, E)}{M(S, E)} \times 100
\end{equation}
  
In which $I_M$ represents the improvement in similarity $M$. 
The condition for a successful attack is defined as:
\[
    \forall s \in S, \forall M \in \mathcal{M}: I_M(s) > P
\]
where $P$ is a predefined threshold. That is a backdoor attack on a summary $s$ is considered successful if and only if the improvement in all three similarity metrics exceeds $P$.

Definition of indicator function:
\begin{equation}   
    \mathbb{I}(s) = 
    \begin{cases} 
        1, & \text{if } \forall M \in \mathcal{M}: I_M(s) > P \\
        0, & \text{otherwise}
    \end{cases}
\end{equation}
Then the ASR is defined as:
\begin{equation}
\label{eq:asr}
    ASR = \frac{1}{n} \sum_{i=1}^n \mathbb{I}(s_i)
\end{equation}
In the backdoor attack, task hinted in code summarizations, upon successful attack, we expect to alter the summarization intended to describe code security into one that describes the code functionality. ASR represents the percentage of test instances where the intent was successfully altered.

\section{Data Application}
Our dataset can be applied to numerous vulnerability-related research areas, fully leveraging its unique advantages, especially the multi-perspective code summaries and security intent analysis. Through these applications, our dataset can bring new perspectives and methods to software security research and practice, driving innovation and development in the field of code security. Specific application scenarios include:
\begin{enumerate}

    \item \textbf{Multidimensional Vulnerability Feature Analysis} Our dataset includes not only traditional CVE information and code changes but also provides functional descriptions and security comments. This multi-perspective data can be used for a more comprehensive vulnerability feature analysis. Researchers can utilize natural language processing (NLP) techniques to perform text mining on CVE summaries, functional descriptions, and security comments, extracting keywords and semantic features. At the same time, static analysis can be performed on code changes to extract structural and syntactic features of the code. By combining these multidimensional features, researchers can develop more comprehensive vulnerability models, gaining deeper insights into the intrinsic characteristics of vulnerabilities and potential security risks. This analysis can help developers better understand and prevent common security vulnerabilities, as well as provide valuable case materials for security training.
    
    \item \textbf{Security Intent-Aware Vulnerability Detection} Traditional vulnerability detection methods mainly focus on the syntactic and structural features of the code, often overlooking the developer's security intent. By using the security intent summaries in our dataset, researchers can develop a new generation of "security intent-aware" vulnerability detection tools. These tools not only analyze the code itself but also take into account the developer's security considerations. For example, a deep learning model can be trained to combine code features with security intent descriptions to identify potential vulnerabilities in the code. This approach can improve the accuracy of vulnerability detection, especially when detecting hidden vulnerabilities resulting from insufficient security awareness among developers.
    
    \item \textbf{Intelligent Vulnerability Repair Recommendation System} By utilizing the code changes and corresponding security intent summaries in our dataset, researchers can develop an intelligent vulnerability repair recommendation system. This system can analyze known vulnerability repair patterns and combine them with the specific security intent of a vulnerability to provide developers with customized repair suggestions. For example, when the system detects a potential buffer overflow vulnerability, it will not only offer the standard repair code but also explain, based on the security intent summary, why this repair is necessary and how to avoid similar issues in future coding. This system can significantly improve the efficiency and quality of vulnerability repairs while enhancing developers' secure coding skills.

    \item \textbf{Trade-off Analysis between Security and Functionality} Our dataset provides both functional descriptions and security intent summaries, offering a unique perspective on the trade-offs between security and functionality in software development. Researchers can analyze how developers make choices between functionality and security considerations across different types of projects or stages of development. By comparing functional descriptions and security intent summaries, it is possible to identify which types of functionalities are more likely to introduce security vulnerabilities, thus guiding software architecture design and development process optimization. This analysis can help project managers and architects make better security decisions early in the project.
    
    \item \textbf{Automated Security Documentation Generation} Using the multidimensional information in our dataset, researchers can develop automated security documentation generation tools. This tool can analyze code, functional descriptions, and security intent summaries to automatically generate detailed security-related documentation. For example, it can generate security considerations for each module or function, describing potential risk points and recommended security practices. This not only improves the quality and consistency of documentation but also ensures that security considerations receive continuous attention throughout the development process. For large projects or fields requiring strict security audits (such as finance and healthcare), this tool can significantly enhance the efficiency of security management.

    \item \textbf{Alignment of Vulnerability Reports and Security Comments} Security practitioners primarily obtain vulnerability information from reports published by sources like CVE and NVD. However, the accuracy and credibility of these reports have come under scrutiny in recent years, with discrepancies often found in details such as attack vectors, affected components, and severity assessments. To address these concerns, our dataset provides a resource for aligning and cross-referencing these reports with actual vulnerable code samples and detailed security comments. This enables a deeper validation process, allowing users to assess whether specific information—such as attack methods, vulnerability triggers, or the impact described in reports—aligns with the code’s actual behavior and risk. Researchers and security analysts can use this alignment to improve the reliability of vulnerability databases, refine security practices, and potentially identify inconsistencies or inaccuracies in public vulnerability reporting.
    
    \item \textbf{Cross-Language Vulnerability Pattern Migration} Although our dataset is primarily based on specific programming languages, the security intent summaries provide a language-independent description of vulnerabilities. Researchers can leverage this feature to study how to transfer vulnerability patterns identified in one programming language to another. For instance, by analyzing the security intent summaries of a certain type of vulnerability in C++, researchers can explore how to identify and prevent similar security issues in Java or Python. This cross-language vulnerability pattern migration research can help develop general security solutions for multilingual environments, which is particularly valuable for large projects that use multiple programming languages.
\end{enumerate}

\section{Limitation}
In the multi-perspective code intent summary dataset we have constructed, there are some limitations and shortcomings:
\begin{enumerate}
   \item \textbf{Language Coverage:} The current version of the dataset focuses primarily on a few mainstream programming languages, such as C/C++ and Python. Our coverage may not be comprehensive for some emerging or less commonly used programming languages, which might limit the dataset's applicability in certain specific domains or technology stacks. We are working to expand the dataset to include a wider variety of programming languages.

    \item \textbf{Timeliness:} Given the rapid evolution of software development and security vulnerabilities, our dataset may not fully reflect the latest security threats and types of vulnerabilities. This is particularly true for some newly emerging vulnerability types or attack methods, where our dataset may lag. We plan to establish a regular update mechanism to ensure the dataset includes the latest security vulnerability information.
    
    \item \textbf{Accuracy of Automated Labeling:} Although we used advanced LLMs for code to perform automatic labeling and conducted manual validation, there may still be some labeling errors or inconsistencies. Especially in the explanation of security intent, some subjectivity may exist. We are developing more stringent quality control processes to improve labeling accuracy and consistency.
    
    \item \textbf{Privacy and Sensitive Information:} While collecting and processing code from open-source projects, we have made every effort to remove all potentially personal or sensitive information. However, some omissions may still exist. We recommend users exercise additional caution when using the dataset and adhere to relevant ethical and legal guidelines.
    
    \item \textbf{Limitations of Contextual Information:} Although we provide code samples, functional descriptions, and security intent summaries, this may not be sufficient to fully understand certain complex vulnerabilities. For instance, some vulnerabilities may require more extensive system architecture knowledge or specific runtime environment information to be fully understood and reproduced.

\end{enumerate}
We recognize these limitations and are actively working to improve and expand the dataset in future versions. We also welcome feedback and contributions from the research community to continuously enhance the quality and utility of the dataset. As new information and improvements are incorporated, we will update the dataset and release new versions regularly.

\section{Conclusion}
This study introduces an innovative multi-perspective code intent summary dataset, a significant contribution to the field of software security and vulnerability analysis. Our dataset includes not only traditional vulnerability information, such as vulnerable source code, but also incorporates unique functional descriptions and security intent summaries. This multidimensional data structure provides researchers and developers with unprecedented insights into the essence of vulnerabilities. Key features of the dataset include comprehensiveness (covering multiple programming languages), multi-perspective analysis (combining code changes, functional descriptions, and security intent summaries), high-quality data (through rigorous selection and validation), practicality (including code versions before and after vulnerability fixes), and innovation (introducing security intent summaries as a new dimension). The dataset has a wide range of applications, benefiting areas from in-depth vulnerability feature analysis to intelligent security tool development and cross-language vulnerability pattern research. It not only promotes the development of academic research but also provides valuable references for real-world software development and security practices.

Looking forward, we plan to further improve and expand this dataset in several areas. First, we aim to broaden language coverage by incorporating samples from more programming languages, especially emerging languages. Second, we will establish a regular update mechanism to ensure that the dataset reflects the latest security threats and vulnerability types. We will also optimize the automatic labeling algorithm to improve the accuracy and consistency of security intent summaries. To better support research on complex vulnerabilities, we plan to provide more detailed system architecture and runtime environment information. Finally, we intend to expand data sources by exploring additional issue tracking and source code management systems, such as Mercurial, Subversion, JIRA, and Bugzilla, beyond the currently used version control systems, to further enrich our dataset.

We believe that with these improvements, our dataset will become an essential resource for software security research and practice, driving innovation and advancement in the field. We also welcome feedback and contributions from the research community as we work together to enhance the security and reliability of software systems.

\end{document}